\documentclass[12pt,preprint]{aastex}
\usepackage{graphicx}
\shorttitle{Flux Calibrator Polarization Characteristics}
\shortauthors{Perley and Butler}
\begin{document}
\title{Integrated Polarization Properties of 3C48, 3C138, 3C147, and 3C286}
\author{R. A. Perley and B. J. Butler}
\email{RPerley@nrao.edu, BButler@nrao.edu}
\affil{National Radio Astronomy Observatory\footnote{The National
    Radio Astronomy Observatory is a facility of the National Science
    Foundation operated under cooperative agreement by Associated
    Universities, Inc.}}
\affil{P.O.Box O, Socorro, NM, 87801}
\slugcomment{Intended for the Astrophysical Journal, Supplement Series}

\begin{abstract}
We present the integrated polarization properties of the four compact
radio sources 3C48, 3C138, 3C147 and 3C286, from 1 to 50 GHz, over a
30-year time frame spanning 1982 to 2012.  These four sources are
commonly used as flux density and polarization calibrators for cm-wave
interferometers.  Using the polarized emission of Mars, we have
determined that the true position angle of the linearly polarized
emission of 3C286 rises from 33 degrees at 8 GHz to 36 degrees at 45
GHz.  There is no evidence for a change in the position angle over
time.  Using these values, the position angles of the intergrated
polarized emission from the other sources are determined as a function
of frequency and time.  The fractional polarization of 3C286 is found
to be slowly rising, at all frequencies, at a rate of $\sim
0.015$\%/year.  The fractional polarizations of 3C48, 3C138, and 3C147
are all slowly variable, with the variations clearly correlated with
changes in the total flux densities of these sources.
\end{abstract}


\maketitle

\section{Introduction}

\citet{PB13} have proposed an absolute spectral flux density scale
valid from 1 to 50 GHz, based on the known absolute emission spectrum
from the planet Mars (\citet{Wei11}), and employing accurate flux
density ratios derived by the VLA approximately yearly since 1983.
They propose that the quasar 3C286, whose flux density has been stable
to $\sim$1\% over the past 30 years, be employed as the primary flux
density calibrator source for frequencies between 1 through 50 GHz,
and provide a polynomial expression for its spectral flux density.  In
addition, \citet{PB13} provide time-variable polynomial expressions
for the spectral flux densities of 3C48, 3C138, and 3C147.

The VLA, prior to its upgrade, always provided full polarization
information when used in its standard continuum operating modes.  The
nineteen separate observing sessions utilized by \citet{PB13} provides
a detailed database from which the polarization properties of the
fourteen target sources can be determined.  In addition, since the
planet Mars was extensively observed in this program, its well known
polarization characteristics can be utilized to establish the true
position angle of polarized emission of the other sources.

In this paper, we briefly describe the polarization calibration
procedure, and give results for the integrated polarization properties
of the four most compact sources in the program -- 3C48, 3C138, 3C147,
and 3C286.

\section{Polarization Calibration}

\subsection{Observing Methodology and Calibration}

The observing methodology and calibration procedures are described in
detail in \citet{PB13}, and will not be repeated here.  The additional
calibration steps employed to determine the antenna polarization, and
the position angle of linearly polarized flux density are given below.

\begin{itemize}
\item The position angle rotation introduced by the earth's ionosphere
  was estimated by utilizing TEC data from the CDDIS data archive with
  a model of the earth's magnetic field. The predicted corresponding
  phase rotation was applied to the LCP data.  The correction is
  typically a few degrees at 1.5 GHz, and is unimportant above $\sim$5
  GHz.
  \item The antenna polarizations were measured utilizing the
  technique of \citet{Con69} which utilizes the rotation of the
  antenna parallactic angle, $\Psi$, over time.  This rotation permits
  a separation of the antenna from the source polarizations, and
  provides estimates of each.
\item The plane of linear polarization was based initially by
  adjusting the phase of the LCP channel so the observed position
  angle of the polarized flux density from 3C286 was 33 degrees at all
  frequencies.
\end{itemize}
Following this calibration procedure, images for all sources in Stokes
parameters Q and U were made for each epoch.  The polarized images of
the optically thin unpolarized source NGC7027 demonstrate the accurcy
of our determinations of the fractional polarization to be typically
0.1\%.

Our observing program comprised fourteen sources, observed at all VLA
frequency bands in the lowest-resolution `D' configuration.  The
polarization imaging results for the heavily resolved sources 3C123,
3C196, and 3C295 show a complicated polarization pattern, with very
low integrated polarization.  Because of this, these sources are not
useful as polarization calibrators.  No linear polarization, to a
limit $\sim$ 0.1\%, was found for the planetary nebulae NGC7027 and
NGC6572, the evolved star MWC349, and the planets Uranus and Neptune.
Venus does show some polarization from its surface at low frequencies,
but the quality of the data and the resolution of the images are both
too low for these results to be useful.

The sources 3C48, 3C138, 3C147, and 3C286 are all significantly
polarized at most frequencies utilized by the VLA.  All are very
compact, and are thus useful for polarization calibration purposes,
providing their integrated polarization properties are at most slowly
variable.  The planet Mars was sufficiently resolved on twelve
observing sessions at our higher frequencies that its polarization can
be imaged and utilized to calibrate the position angle for the other
sources.  The procedure is described in the following section.

\section{The True Position Angle for the Polarized Flux of 3C286}

Most VLA users have adopted the value of 33 degrees for the position
angle of the linearly polarized flux density of 3C286 at all frequencies.
This value was determined by \citet{Big73} with a quoted error of $\pm
0.9$ degrees, and is based on observations at 6.7 GHz with the
Algonquin Radio Telescope.  \citet{Big73} also give a position angle
of 31$\pm1.3$ degrees for 3C286 at 10.7 GHz.  We are not aware of any
systematic effort to measure the true position angle of 3C286 at other
frequencies in order to check the common assumption that the position
angle is independent of wavelength.  Our observations of Mars provide
the capability to make this check for the higher frequencies where we
are able to resolve the planet's disk, as the emission of blackbody
radiation from a planet such as Mars will show a radially oriented
linear polarization which maximizes near the limb of the planet
(\citet{Hei63}).

For a solid body such as Mars, the thermal radio emission originates
from beneath the surface, which upon crossing the surface, is refracted
according to Snell's law
\begin{equation}
\sin\theta_t = \frac{n_i}{n_t}\sin\theta_i,
\end{equation}
where $\theta$ is the propagation angle with respect to the surface
normal, $n=\sqrt{\epsilon}$ is the index of refraction, $\epsilon$ is
the dielectric constant for the particular medium, and the subscripts
$i$ and $t$ represent the media for the incident and transmitted
radiation.  (We assume non-magnetic media, for which $\mu = 1$).
Because of the change in media properties at the planetary surface,
from $n_i > 1$ to $n_t\sim 1$, part of the emerging wave is
reflected, and part transmitted.  The fraction transmitted depends
upon the angle of incidence, the dielectric constant, and the
polarization.  The transmissivity (defined as the fraction of the
incident power which is transmitted across the boundary) is given by
\citet{BW80}
\begin{eqnarray}
T_\parallel &=& \frac{\sin 2\theta_i \sin 2\theta_t}
              {\sin^2(\theta_i+\theta_t)\cos^2(\theta_i-\theta_t)}\\
T_\perp     &=& \frac{\sin 2\theta_i \sin 2\theta_t}
              {\sin^2(\theta_i+\theta_t)}.
\end{eqnarray}
The parallel linear polarization component lies in the plane defined
by the incident and transmitted propagation directions.  The
perpendicular component is orthogonal to this plane.  

Assuming a simple model where Mars is a uniform sphere of radius $R$,
the radiation emitted towards the observer from a location with
perpendicular offset $d$ from the center of the visible disk leaves
the planet surface at angle $\sin(\theta)=d/R$ to the local planet
normal.  Defining $x = d/R$ as the fractional perpendicular offset
from the planet center, the transmissivity can be written, using
equations (1), (2), and (3) as
\begin{eqnarray}
T_\parallel(x) &=& \frac{4\epsilon\sqrt{1-x^2}\sqrt{\epsilon-x^2}}
                 {\left(\epsilon\sqrt{1-x^2}+\sqrt{\epsilon-x^2}\right)^2}\\
T_\perp(x)     &=& \frac{4\sqrt{1-x^2}\sqrt{\epsilon-x^2}}
                 {\left(\sqrt{1-x^2}+\sqrt{\epsilon-x^2}\right)^2}.
\end{eqnarray}
The parallel component is radially oriented, and the perpendicular
component is azimuthally oriented.  It can be shown that $T_\parallel
> T_\perp$ for $0 < x < 1$.  Further, $T_\parallel$ rises from the
planet center outwards, reaches a maximum of unity at the Brewster
angle, $x = \sqrt{\epsilon/(1+\epsilon)}$, and falls dramatically
thereafter, while $T_\perp$ declines uniformly with offset.  Hence,
the observed emission will be radially polarized, with the fractional
polarization increasing with offset.  The fractional degree of
polarization 
\begin{equation}
p = \frac{T_\parallel(x)-T_\perp(x)}{T_\parallel(x)+T_\perp(x)}
\end{equation}
is a strong function of the planet dielectric constant $\epsilon$,
rising from zero at the planet center to a maximum very near to the
planet's limb (\citet{Hei63}).  For Mars, with $\epsilon\sim 2.5$
(\citet{Rudy87}), the maximum linear polarization would exceed 30\%,
were the Martian surface smooth.

Finite angular resolution and surface roughness will reduce the
observed fractional polarization (\citet{Hagfors65},
\citet{Alekseev68}, \citet{Golden79}), but the radial position angle
relation will be preserved on large scales, allowing a position angle
calibration to be established for observations which adequately
resolve the planet's disk.

\begin{figure}[ht]
\centerline{\hbox{
\includegraphics[width=3in,angle=-90,origin=c]{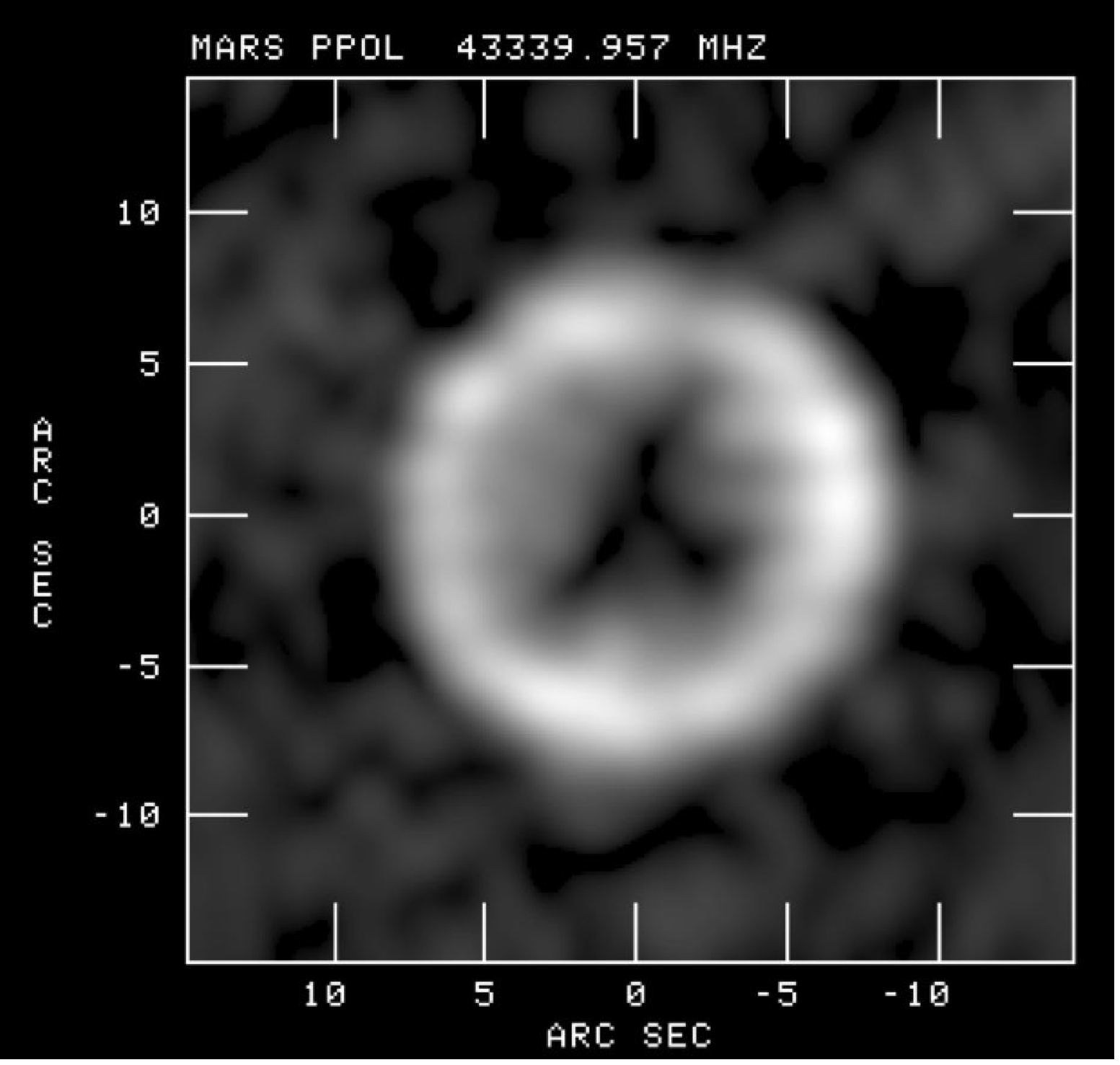}
\includegraphics[width=3in]{QPol-mod.eps}}}
\centerline{\parbox{6in}{
\caption{\small (Left) A grey-scale image of the linearly polarized
  intensity from Mars, with resolution 2 arcseconds, on 16 April,
  1999.  The planet diameter is 15.6 arcseconds.  (Right) A contour
  plot of the same image, with the apparent position angle of
  polarized flux density superposed.  A small deviation from purely radial is
  apparent.}
\label{fig:QMars99}}}
\end{figure}

To illustrate this effect, we show in Fig.~\ref{fig:QMars99} images of
the polarized intensity, and the position angle of the polarized
intensity at 43 GHz, taken from our observations on April 16, 1999,
when Mars had an angular diameter of 15.6 arcseconds. The instrumental
resolution is 2.0 arcseconds.

Twelve of the thirteen observing sessions which included the planet
Mars were taken when Mars had an angular diameter sufficiently large
to resolve its polarization in at least one frequency band.  For these
sessions, we made polarization images of Mars at all frequencies where
there were at least three resolution elements across the planet.  The
lowest frequency at which this could be done for our observations was
5 GHz.  A special {\tt AIPS} task was written to find the mean
deviation of the observed position angle around the planet's disk from
the radial direction.  This deviation was then added to the assumed
value of 33 degrees for 3C286 to determine the correct position angle
for that source.  The measured position angles for the remaining
sources were subsequently corrected using this offset.  The observed
deviations, and (1-$\sigma$) error are given in
Table~\ref{tab:PAngle}.  The bottom line in Table~\ref{tab:PAngle}
gives the weighted mean of the data and the probable error.  A small
but significant offset is found.  The corrected values for 3C286 are
given in Table~\ref{tab:3C286PA}.

\begin{deluxetable}{cccccc}
\tabletypesize{\scriptsize} 
\tablecaption{Offset Angles from
     $33^\circ$ of Mars' Polarized Emission
\label{tab:PAngle}}
\tablewidth{0pt}
\tablehead{\colhead{Session}&\colhead{3 cm}&
\colhead{2 cm}& \colhead{1.3 cm}& \colhead{1.0 cm}&
\colhead{0.7 cm}} 
\startdata 
1995&            &            &-2.5$\pm$0.6&            &            \\
1999&-1.0$\pm1.8$&-0.5$\pm1.0$&-2.9$\pm$1.0&            &-4.4$\pm$0.5\\
2000&            &            &            &            &-3.5$\pm$5.1\\
2001&            &-0.7$\pm7.0$&-3.1$\pm$3.2&            &-2.2$\pm$4.3\\
2003&            &            &-0.4$\pm$4.0&            &-3.7$\pm$2.5\\
2004&            &            &            &            & 2.3$\pm$3.5\\
2006&            &-0.3$\pm2.4$&-3.2$\pm$1.1&            &-0.5$\pm$1.2\\
2007&            &            & 2.0$\pm$8.8&            & 0.4$\pm$3.7\\
2008&            &            &            &            & 0.5$\pm$3.8\\
2010&-0.6$\pm4.3$&            &-1.5$\pm$1.2&-2.6$\pm1.1$&-4.4$\pm$0.5\\
    & 0.7$\pm4.0$&            &-1.3$\pm$1.6& 1.9$\pm1.4$&            \\
2011&            &-1.8$\pm3.2$&-1.6$\pm$2.2& 1.7$\pm2.0$& 0.5$\pm$3.6\\
    &            &            &-1.9$\pm$3.9& 3.9$\pm1.3$&            \\
2012& 0.6$\pm3.9$&-3.5$\pm1.3$& 0.4$\pm$0.8&-4.1$\pm1.0$&-2.6$\pm$0.9\\
    & 0.5$\pm2.9$&-1.2$\pm1.5$&-2.3$\pm$0.8&-4.5$\pm0.8$&-3.3$\pm$0.6\\
\hline
Wgt.Mean&-0.3$\pm$0.9&-1.6$\pm$0.3&-1.91$\pm$0.12&-2.50$\pm$0.33&
-2.77$\pm$0.10\\
\enddata
\tablecomments{The last line gives the weighted mean of the tabular
  entries for each wavelength band.  The listed error is the weighted
  (1-$\sigma$) error.}
\end{deluxetable}

\begin{deluxetable}{cccccc}
\tabletypesize{\scriptsize}
\tablecaption{The True Polarization Position Angle for 3C286
\label{tab:3C286PA}}
\tablewidth{0pt}
\tablehead{\colhead{P.A.}&\colhead{5 GHz}&
\colhead{15 GHz}& \colhead{23 GHz}& \colhead{33 GHz}&
\colhead{45 GHz}} 
\startdata 
3C286&33$\pm$ 1&34.5$\pm$0.5&35$\pm$0.2&35.5$\pm$0.4&35.8$\pm$0.1\\
\enddata
\end{deluxetable}
 
\subsection{Polarization Properties for the Four Primary Calibrators}

\begin{figure}[ht]
\centerline{\hbox{
\includegraphics[width=6in]{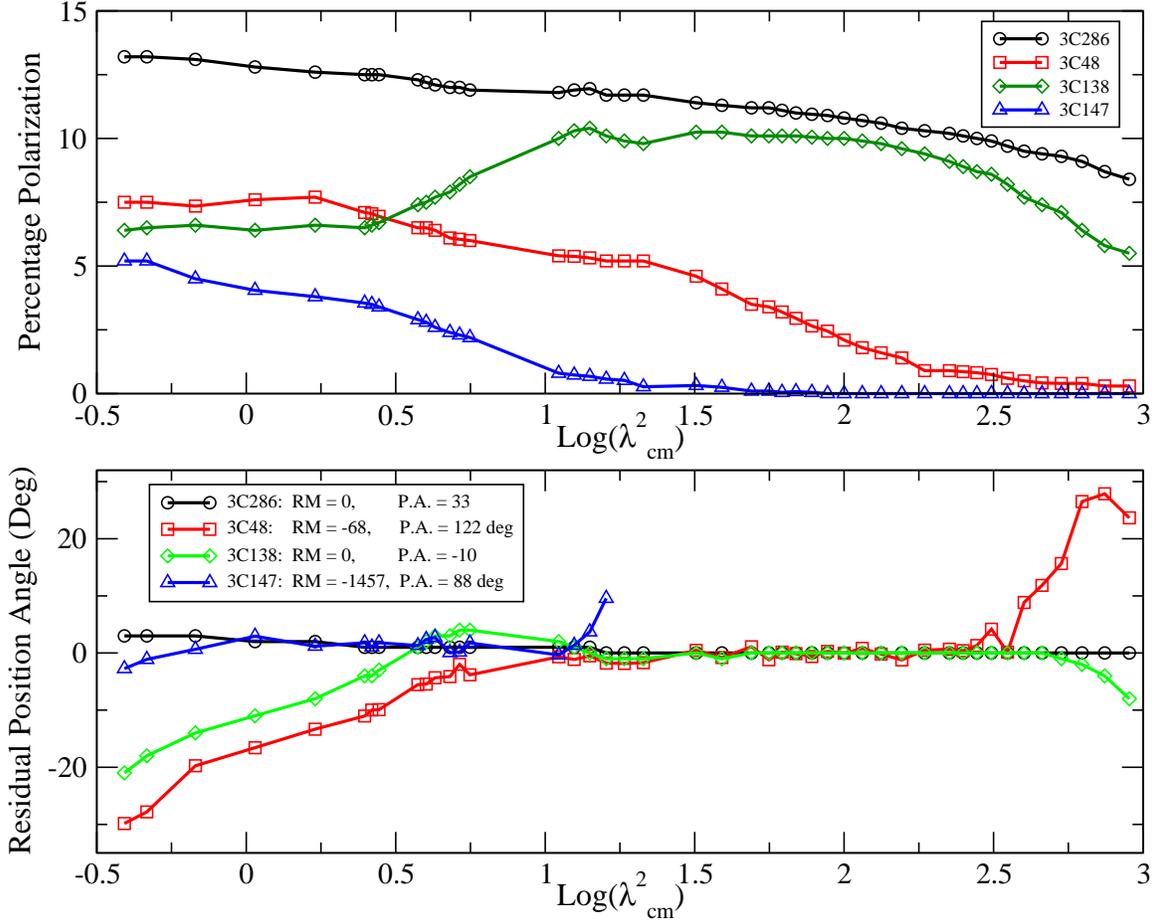}}}
\centerline{\parbox{6in}{
\caption{\small The fractional polarization, and the position angle,
  for the four primary calibrators, as a function of frequency.  These
  data are from the Dec 2010 session.  Note that the RM values shown
  in the lower frame have been used to unwrap the position angles.}
\label{fig:PolPlot}}}
\end{figure}

In Table~\ref{tab:Polarization} we show the fractional linear
polarizations, and the position angles of the polarized intensity from
the four compact sources, takn from the observation run of December
2010.  The position angles are referenced to 3C286, which was set to
33 degrees for frequencies between 1 and 8 GHz, and to the angles
given in Table~\ref{tab:3C286PA} for higher frequencies. These same
data, in graphical form, are shown in Fig.~\ref{fig:PolPlot},
following adjustment for the rotation measure.  In this figure, the
position angle data for 3C147 at frequencies less than 7 GHz are not
shown, as the source is too heavily depolarized to allow a meaningful
determiniation of the position angle.

\begin{deluxetable}{ccccccccc}
\tabletypesize{\scriptsize} 
\tablecaption{Polarization Properties of 3C48, 3C138, 3C147, and 3C286
\label{tab:Polarization}}
\tablewidth{0pt} 
\tablehead{ \colhead{Freq.}& \colhead{3C48p}&
\colhead{3C48$\chi$}& \colhead{3C138p}& \colhead{3C138$\chi$}&
\colhead{3C147p}& \colhead{3C147$\chi$}&\colhead{3C286p}&
\colhead{3C286$\chi$}\\
\colhead{GHz}&\colhead{\%}&\colhead{deg.}&\colhead{\%}&
\colhead{deg.}&\colhead{\%}&\colhead{deg.}&\colhead{\%}&\colhead{deg.}}

\startdata 
1.050&0.3& 25 & 5.6&-14&$<.05$&  --& 8.6&33\\ 
1.450&0.5& 140& 7.5&-11&$<.05$&  --& 9.5&33\\ 
1.640&0.7& -5 &8.4 &-10&$<.04$&  --& 9.9&33\\ 
1.950&0.9&-150& 9.0&-10&$<.04$&  --&10.1&33\\ 
2.450&1.4&-120&10.4& -9&$<.05$&  --&10.5&33\\
2.950&2.0&-100&10.7&-10&$<.05$&  --&10.8&33\\ 
3.250&2.5&-92 &10.0&-10&$<.05$&  --&10.9&33\\ 
3.750&3.2& -84& -- & --&$<.04$&  --&11.1&33\\ 
4.500&3.8& -75&10.0&-11&   0.1&-100&11.3&33\\ 
5.000&4.2&-72 &10.4&-11&   0.3&   0&11.4&33\\ 
6.500&5.2& -68& 9.8&-12&   0.3& -65&11.6&33\\ 
7.250&5.2& -67&10.0&-12&   0.6& -39&11.7&33\\
8.100&5.3& -64&10.4&-10&   0.7& -24&11.9&33\\ 
8.800&5.4& -62&10.1& -8&   0.8& -11&11.9&33\\ 
12.80&6.0& -62& 8.4& -7&   2.2&  43&11.9&34\\
13.70&6.1& -62& 7.9& -7&   2.4&  48&11.9&34\\ 
14.60&6.4& -63& 7.7& -8&   2.7&  53&12.1&34\\ 
15.50&6.4& -64& 7.4& -9&   2.9&  59&12.2&34\\
18.10&6.9& -66& 6.7&-12&   3.4&  67&12.5&34\\ 
19.00&7.1& -67& 6.5&-13&   3.5&  68&12.5&35\\ 
22.40&7.7& -70& 6.7&-16&   3.8&  75&12.6&35\\
23.30&7.8& -70& 6.6&-17&   3.8&  76&12.6&35\\ 
36.50&7.4& -77& 6.6&-24&   4.4&  85&13.1&36\\ 
43.50&7.5& -85& 6.5&-27&   5.2&  86&13.2&36\\ 
\enddata

\tablecomments{Listed are the properties as measured in December 2010.
  The polarization of 3C138 has been changing at the higher
  frequencies due to a large flare which began in 2003.  The other
  sources show relatively stable polarization over time.}
\end{deluxetable}

The position angle of the linearly polarized flux density changes
rapidly with wavelength for 3C48 and 3C147.  For both, the
relationships are well fitted by a $\lambda^2$ law: $\chi(\lambda) =
\mathrm(RM)\lambda^2 + \chi_0$, over part of the frequency range
spanned by our observations.  We made a least squares fit to this
function for all four sources, with the results given in
Table~\ref{tab:RM}.
\begin{deluxetable}{cccc}
\tabletypesize{\scriptsize}
\tablecaption{RM Values for the Four Sources
\label{tab:RM}}
\tablewidth{0pt}
\tablehead{\colhead{Source}&\colhead{Wavelength Range}&
\colhead{RM}& \colhead{$\chi_0$}\\
&\colhead{cm}&\colhead{$\mathrm{rad/m^2}$}&\colhead{deg.}} 
\startdata 
3C48 &1 -- 18   &-68  &122\\
3C138&2 -- 22   & 0   &-10\\
3C147&1 -- 3    &-1467& 88\\
3C286&1 -- $>30$& 0   & 33\\
\enddata
\end{deluxetable}
 
\begin{figure}[ht]
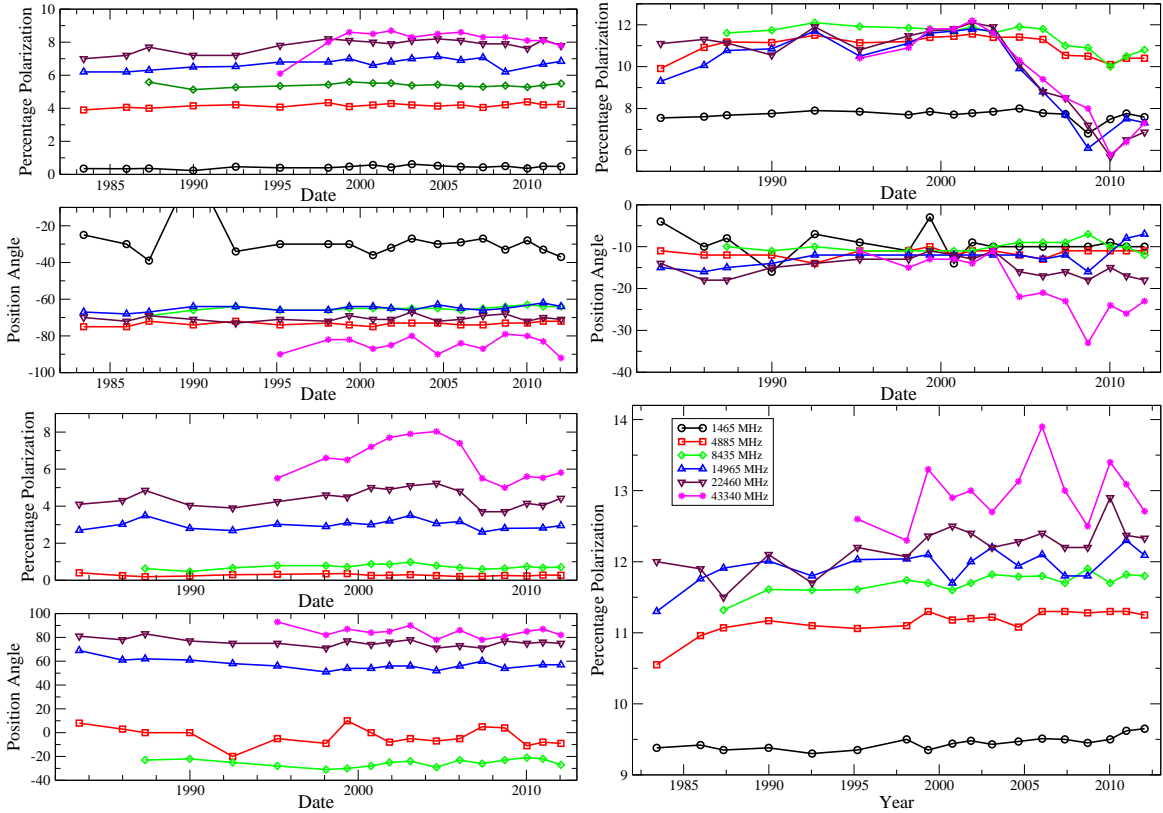

\centerline{\hbox{
\includegraphics[width=3in]{3C48PolFix.eps}
\includegraphics[width=3in]{3C138PolFix.eps}}}
\centerline{\hbox{
\includegraphics[width=3in]{3C147PolFix.eps}
\includegraphics[width=3in]{3C286PolFix.eps}}}
\centerline{\parbox{6in}{
\caption{\small The percentage polarization, and the polarization
  position angle for 3C48 (upper left), 3C138 (upper right), 3C147
  (lower left), and 3C286 (lower right, fractional polarization
  only).}
\label{fig:PolChange}}}
\end{figure}

Of interest is any variation in the polarization properties of these
sources over time.  The recent outburst in 3C138 and the smaller
changes in 3C48 and 3C147 (see \citet{PB13}) may be expected to result
in changes in polarization, as the polarized emission from the
emerging component may not align with that from the larger-scale,
older emission. Shown in Fig.~\ref{fig:PolChange} is the temporal
variation in the fractional polarization and in the position angles of
the four primary calibration sources since 1983.
The figure shows that the fractional polarization of 3C48 has
undergone a small but notable rise since this monitoring program
began, although there appears to be no effect in the position angle.
As expected, there was a dramatic change in the polarization
properties in 3C138 at higher frequencies, beginning in 2003, when the
flare began.  With the flare intensity now waning, the polarization
properties appear to be returning to the same values noted before
2003.  The flux density changes noted for 3C147 are accompanied by
small changes in its polarization, particularly at higher frequencies.
The fractional polarization of 3C286 appears to be slowly, and
steadily rising over the period. A simple linear fit to the data
provides the values shown in Table~\ref{tab:286PolVar}, giving the
fractional polarization in percent for epoch 2000.0, and its change in
percent/century.
\begin{deluxetable}{ccc}
\tabletypesize{\scriptsize}
\tablecaption{Change in Fractional Polarization of 3C286
\label{tab:286PolVar}}
\tablewidth{0pt}
\tablehead{\colhead{Freq.}&\colhead{P(2000)}&\colhead{Slope}\\
\colhead{GHz}& \colhead{\%}& \colhead{\%/century}} 
\startdata 
1.465&$9.450\pm .015$&$0.77\pm 0.17$\\
4.885&$11.18\pm .02$&$1.0\pm 0.2$\\
8.435&$11.67\pm .02$&$1.4\pm 0.3$\\
14.965&$11.97\pm .04$&$0.8\pm 0.5$\\
22.485&$12.20\pm .05$&$2.5\pm 0.6$\\
43.340&$12.87\pm .16$&$2.2\pm 2.3$\\
\enddata
\end{deluxetable}

\section{Discussion}

All four sources have been extensively observed with the VLBA, MERLIN,
and EVN.  Here we briefly compare the polarimetric results from these
instruments against our own.

\subsection{3C48}
The source was the subject of an extensive study by \citet{An10},
untilizing the VLBA, MERLIN, and the EVN.  Their paper shows that the
emission from 3C48 on milliarcsecond scales is quite complicated, with
a short and twisted jet-like structure blending into an extended
region located to the NE of the quasar core.  Our global rotation
measure value of -68 $\mathrm{rad/m}^2$ is in good agreement with the
`spot' values from their component `C' of -68 to -95
$\mathrm{rad/m}^2$.  The MERLIN polarimetric image at 1.65 GHz shows
the weak polarized emission aligned nearly vertically, in good
agreement with our integrated value of -5 degrees at 1.64 GHz.  The
heavily resolved VLBA images shown in their Figure 6 at 4.78 and 8.31
GHz are difficult to compare to our integrated values, but the
polarized position angles shown are close to our integrated values.
High resolution VLA images taken at 26 GHz with 70 mas resolution show
the majority of the polarized emission coming from bright northern
component labelled `C' in \citet{An10}, with the more compact
structures labeled by them at `B', `B2', and `B3'.
   
\subsection{3C138}
The small-scale structure is shown in \citet{Cot97a} to comprise a
relative weak unresolved nuclear core, a strong and well--resolved
jet-like structure extending $\sim$400 milliarcseconds to the NE, and
a very weak counter-jet 250 milliarcseconds to the SW.  The NE jet is
highly and uniformly polarized, and accounts for nearly all the total
polarized emission seen in our integrated measures.  Their Figure 4
shows the weaker, trailing parts of the NE jet to have a linearly
polarized emission at p.a. $\sim$ -25 degrees, while the much brighter
head of the jet to have its polarized emission at an angle near 0.
The sum of the two components is near our integrated measure of
$\sim$-10 degrees.  Our data show the position angle to decrease by
about 15 degrees as the frequency rises -- this is probably due to a
spectral index effect whereby the head of the NE jet becomes more
prominent compared to the steeper spectrum trailing areas.  The
decline in integrated polarization, and decrease in the resulting
position angle, seen since 2002 at the higher frequencies are likely
due to an emerging new component along p.a. near -30 (as shown in
Figure 6 in \citet{Cot97a}) with polarization along this angle.

\subsection{3C147}
VLBA polarimetric imaging at 5 and 8 GHz for 3C147 has been presented
by \citet{Ros09}. At milliarcsecond resolution, the source comprises a
compact core and apparently rapidly expanding jet, extending only
$\sim$10 milliarcseconds to the SW.  Beyond this, a very diffuse and
well-resolved lobe extends to the SW to a maximum extent of $\sim 180$
milliarcseconds.  High resolution VLA observations at $\sim$70
milliarcseconds resolution show another, more extended and detached
elogated lobe extending northwards $\sim$ 700 milliarcseconds.  3C147
has an extremely high RM of -1467 $\mathrm{rad/m}^2$, in good
agreement with the value resported by \citet{Ros09} for their
component B, located at the end of the inner, flaring jet.  High
resolution VLA polarimetric imaging at 25 GHz with 70 milliarcsecond
resolution shows that the nucleus and inner jet regions dominate the
total polarized flux density.  The true integrated position angle of
88 degrees from our integrated measures agrees well with the intrinsic
angle for the B-field of the flaring jet shown by \citet{Ros09} in
their Figure 7.

\subsection{3C286}
 The quasar 3C286 was selected by \citet{PB13} as the primary
non-variable flux density standard from 1 to 50 GHz, based on the
stability of its total intensity over period exceeding 30-years.  This
current study shows that 3C286 also serves as an equally useful and
stable reference source for polarimetry.  Although we detect a very
small secular increase in its polarized flux density, the change
appears to be steady, permitting easy calculation of the expected
polarized flux density at any epoch.

High resolution polarimetry of this source is shown by \citet{Jia96}
and \citet{Cot97b}.  The latter paper sketches the overall structure,
comprising a central core, a compact eastern lobe about 1
arcsecond east, and an extended western lobe, extending $\sim$ 2.5
arcseconds to the SE.  The central core region contains a highly
polarized extension of length $\sim$80 milliarcseconds, in p.a. 135
degrees -- i.e., pointing towards the SE lobe.  Both the total
and polarized emission from this source are unusual.  There is no
clearly defined inverted spectrum core, suggesting either a dearth of
recent activity, or a geometry whereby the core emission is beamed
away from our direction.  The polarized emission is almost entirely
along the jet, contrary to the general rule where jet emission
is primarily polarized transversely to the jet axis.  

3C286 has recently been proposed by \citet{Agu12} as a polarization
calibration source for millimeter-wavelength observations.  Their
results are in excellent agreement with a simple extrapolation of
ours: a polarization fraction of 13.5\% and an intrinsic position
angle of 37.4 degrees at $\lambda=3$ mm.  \citet{Agu12} also provide
values at $\lambda=1$ mm of 14.5\% and 33.5 degrees -- however, these
have much larger error estimates, and are not inconsistent with our
results.

\section{Summary}

We have determined that the position angle of linearly polarized
emission from 3C286 rises from 33 degrees at frequencies below 10 GHz,
to 36 degrees at 43 GHz.  Changes in this position angle over time are
less than $\sim$2 degrees over the past 20 years.  The fractional
linear polarization of 3C286 is steadily increasing at all frequencies
at a rate of about 0.15\%/year.  

The polarization characteristics of 3C48 and 3C147 are fairly stable,
but small changes, likely related to the small variations in total
intensity noted by \citet{PB13} are visible. The polarization
position angles for these two objects for frequencies at which their
fractional polarization exceed 1\% are well fitted with a simple
$\lambda^2$ law over most of the observed frequency range.    

The strongly polarized source 3C138 has undergone a notable flare,
beginning in 2003, which nearly doubled its total intensity by 2010,
and has been rapidly declining since.  The polarization properties
also changed dramatically during this period, and appear to be
returning to the pre-flare levels.  

\section{Acknowledgement}

We thank Eric Greisen for generating the special AIPS task to
calculate the mean offset in polarization position angle.

\end{document}